\documentclass[11pt]{pazh}

\usepackage{psfig}

\begin{document}
	
\setcounter{page}{429}

\sloppypar

\title{\bf The structure and evolution of M\,51-type galaxies}

\author{\copyright\,2003\,\,\, V.P.Reshetnikov, S.A.Klimanov}

\institute{Astronomical Institute of St.Petersburg State University,
Universitetskii pr. 28, Petrodvoretz, 198504 Russia}

\authorrunning{Reshetnikov, Klimanov}
\titlerunning{M\,51 type galaxies}

\abstract{We discuss the integrated kinematic parameters of 20 M\,51-type
binary galaxies. A comparison of the orbital masses of the galaxies with 
the sum of the individual masses suggests that moderately
massive dark halos surround bright spiral galaxies. The relative velocities
of the galaxies in binary systems were found to decrease with increasing 
relative luminosity of the satellite. We obtained evidence that the
Tully--Fisher relation for binary members could be flatter than that for
local field galaxies. An enhanced star formation rate in the binary 
members may be responsible for this effect. In most binary systems,
the direction of orbital motion of the satellite coincides with the
direction of rotation of the main galaxy. Seven candidates for distant 
M\,51-type objects were found in the Northern and Southern Hubble Deep
Fields. A comparison of this number with the statistics of nearby galaxies
provides evidence for the rapid evolution of the space density of 
M\,51-type galaxies with redshift $z$. We assume that M\,51-type
binary systems could be formed through the capture of a satellite by a
massive spiral galaxy. It is also possible that the main galaxy and its 
satellite in some of the systems have a common cosmological origin.
}
\titlerunning{M\,51-type galaxies}
\maketitle

\section{Introduction}

M\,51-type (NGC 5194/95) binary systems consist of a spiral galaxy 
and a satellite located near the end of the spiral arm of the main 
component. Previously (Klimanov and Reshetnikov 2001), we analyzed
the optical images of about 150 objects that were classified by 
Vorontsov-Vel'yaminov (1962--1968) as galaxies of this type. Based 
on this analysis, we made a sample of 32 objects that are most likely 
to be M\,51-type galaxies. By analyzing the sample galaxies, we
were able to formulate empirical criteria for classifying
a binary system as an object of this type (Klimanov and Reshetnikov 2001): 

(1) the $B$-band luminosity ratio of the components ranges from 1/30 to 1/3; 

(2) the satellite lies at a projected distance that does not exceed two
optical diameters of the main component.

Klimanov et al. (2002) presented the results of spectroscopic observations 
of 12 galaxies from our list of M\,51-type objects, including rotation curves
for the main galaxies and line-of-sight velocities for their satellites. 
Together with previously published results of other authors, we have 
access to kinematic data for most (20 objects) of our selected 
M\,51-type galaxies. Here, we analyze the observational data
for the galaxies of our sample. All of the distance-dependent quantities 
were determined by using the Hubble constant 
H$_0$ = 75 km s$^{-1}$ Mpc$^{-1}$ (except Section 4).

\section{Mean parameters of the galaxy sample}

\begin{figure*}[!ht]
\centerline{\psfig{file=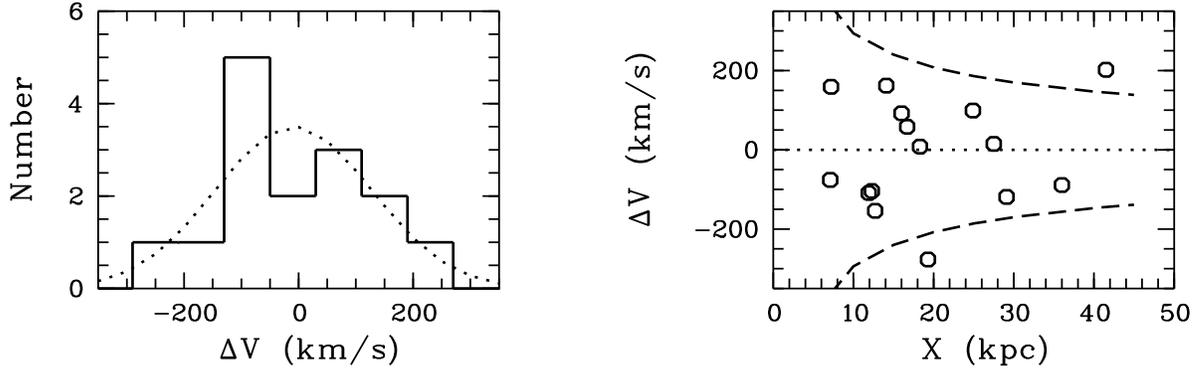,width=16cm,angle=-90,clip=}}
\caption{{\it Left}: The distribution of observed line-of-sight velocity 
differences between the main galaxy and its satellite. {\it Right}: The
velocity difference between the main galaxy and its satellite versus the
projected linear separation between them.}
\end{figure*}

The kinematic parameters for 12 binary systems were published 
by Klimanov et al. (2002) (see Fig. 1 and the table in this paper). 
The corresponding data for eight more systems from our list 
(object nos. 6, 10, 13, 14, 19, 21, 25, and 27; see Table 7 and Fig. 7
in Klimanov and Reshetnikov 2001) were taken from the LEDA and NED databases.

The mean $B$-band absolute magnitude of the main galaxy for 20 binary 
systems is -20\fm6$\pm$0\fm3 (below, the standard deviation from the mean 
is given as the error). Therefore, the main components are relatively
bright galaxies comparable in luminosity to the Milky Way. The mean 
luminosity ratio of the satellite and the main galaxy is 0.16$\pm$0.04. 
The mean apparent flattening of the main galaxy 
($\langle b/a \rangle$=0.61$\pm$0.04) and its satellite (0.66$\pm$0.03) 
roughly correspond to the expected flattening of a randomly 
oriented thin disk (2/$\pi$=0.64). The satellites lie at a
projected distance of  $\langle X \rangle$=19.0$\pm$2.7 kpc or, in 
fractions of the standard optical radius R$_{25}$ measured
from the $\mu(B)$=25$^m$/arcsec$^2$ isophote, at a distance of 
(1.39$\pm$0.11)R$_{25}$. In going from the projected linear distance 
to the true separation, we find that the satellites are separated 
by a mean distance of  $\approx$4/$\pi$\,$\cdot$\,1.39 R$_{25}$ = 
1.8 R$_{25}$.

The mean maximum rotation velocity of the main galaxy corrected 
for the inclination of the galactic plane and for the deviation of 
the spectrograph slit position from the major axis (see the next section
for more details) is $\langle V_{max} \rangle$=190$\pm$19 km/s. If we
exclude the galaxies with $b/a>0.7$ seen almost face-on, then this 
value increases to 203$\pm$16 km/s (12 pairs).

The mean difference between the line-of-sight velocities of the galaxies 
and their satellites is $\langle \Delta V \rangle$=--9$\pm$35 km/s.
The absolute value of this difference, i.e., the difference with no 
sign, is 115$\pm$18 km/s. The relatively low values of $\Delta V$ 
suggest that all objects of our sample are physical pairs. Remarkably,
the mean absolute value of the velocity difference for M\,51-type 
galaxies is close to that (137$\pm$6 km/s) for 487 galaxy pairs from 
the catalog of Karachentsev (1987) with a ratio of the orbital mass 
of the pair to the total luminosity of its components $f<100$. Note
also that the measurement error of $\Delta V$ in Klimanov et al. (2002) 
is, on average, only 20 km/s.

Figure 1 (left panel) shows the observed $\Delta V$ distribution. 
To a first approximation, this distribution can be described by a 
Gaussian with the standard deviation  $\sigma$=140 km/s (dotted line). 
Right panel of Fig. 1 shows a plot of the difference $\Delta V$ against 
the projected linear separation between the main galaxy and its
satellite. The dashed lines in the figure represent the expected 
(Keplerian) dependences for a point mass with 
M$_{tot}$=2$\times$10$^{11}$\,M$_{\odot}$ located at the center of the
system.
 
Let us compare the mean values of the individual and orbitalmass 
estimates for M\,51-type galaxies. For a sample of binary galaxies 
with the plane of the circular orbit randomly oriented relative to the
line of sight, the total mass of the components can be found as 
M$_{orb}$=(32/3$\pi$)(X$\Delta V^2/G$), where $G$ is the gravitational
constant (Karachentsev 1987). For the objects of our sample,  
$\langle$M$_{orb}$$\rangle$=(2.9$\pm$1.0) $\times$ 10$^{11}$\,M$_{\odot}$. 
We will determine the individual masses of the main components by using 
the maximum rotation velocities (see the next section) and by assuming that
the rotation curves of the galaxies within their optical radii are flat. 
For a spherical distribution of matter, we then obtain the mean mass of 
the main galaxy, 
$\langle$M$_{main}$$\rangle$=(1.6$\pm$0.4) $\times$ 10$^{11}$\,M$_{\odot}$, 
and the mass-to-luminosity ratio,  
$\langle$M$_{main}/L_{main}(B)\rangle$=4.7$\pm$0.9 M$_{\odot}$/$L_{\odot,B}$.
The ratio of the orbital mass of M\,51-type systems to the mass of the 
main galaxy for the objects under consideration is 1.9$\pm$0.5. 
Given the satellite's mass, the ratio of the orbital mass of the system 
to the total mass of the two galaxies is $\approx$1.6 (for a fixed 
mass-to-luminosity ratio and a mean luminosity ratio of the satellite 
and the main galaxy equal to 0.16). If the orbits of the satellites are 
assumed to be not circular but elliptical with the mean eccentricity 
$e$=0.7 (Ghigna et al. 1998), then the orbital mass estimate increases
by a factor of 1.5 (Karachentsev 1987). The ratio of the orbital mass 
of the system to the total mass of the two galaxies also increases 
by a factor of 1.5 (to 2.4). Consequently, we obtained evidence for 
the existence of moderately massive dark halos around bright spiral
galaxies within (1.5--2)R$_{25}$.

In Fig. 2, $k= |\Delta V|/V_{max}$ is plotted against the ratio of the 
observed $B$-band luminosities of the satellite and the main galaxy 
($L_s/L_m$), where $V_{max}$ is the maximum rotation velocity of the 
main galaxy. If $|\Delta V|/V_{max}$ is close to unity, then the relative 
velocity of the satellite is approximately equal to the  disk rotation 
velocity of the main galaxy; if $k\approx$0, then the satellite's observed 
velocity is close to the velocity of the main galaxy. We see a clear 
trend in the figure -- relatively more massive satellites show lower
values of $k$. If we restrict our analysis to satellites with 
$L_s/L_m < 0.5$ (circles), then the correlation coefficient
of the dependence shown in Fig. 2 is -0.77; i.e., the correlation is 
statistically significant at $P>$99\% (the diamond in the figure 
indicates the parameters of the system NGC\,3808A,B with $L_s/L_m$=0.64.)
The corresponding linear fit is indicated in the figure by the dashed 
straight line. Low-mass satellites with $L_s/L_m<0.2$ are located, 
on average, on the extension of the rotation curve for the main 
galaxy and have velocities close to $V_{max}$ (for them,  
$\langle k \rangle = 1.13\pm0.19$).  More massive satellites with 
$0.2 < L_s/L_m < 0.5$ have a lower relative velocity:  
$\langle k \rangle = 0.52\pm0.15$.

\begin{figure}
\centerline{\psfig{file=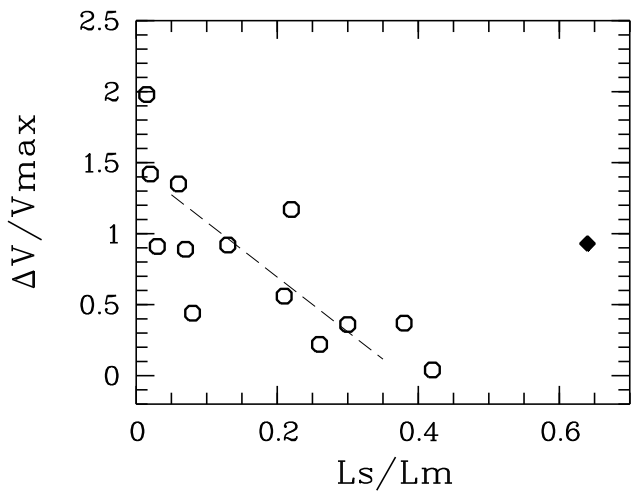,width=8.8cm,angle=-90,clip=}}
\caption{The relation between the satellite's relative orbital
velocity and the luminosity ratio of the satellite and the
main galaxy.}
\end{figure}

There is another curious trend: more distant satellites are, on average, 
more massive. This trend can be explained by observational selection 
when choosing candidates for M\,51-type galaxies. Another plausible 
explanation is that the massive satellites located near the main galaxy 
have a shorter evolution scale, because the characteristic lifetime 
of a satellite near a massive galaxy due to dynamical friction is 
inversely proportional to the satellite's mass (Binney and Tremaine 1987).

The optical radius of the satellite is plotted against the projected 
linear distance to the center of the main galaxy in Fig. 3. The dashed 
straight line in the figure indicates the expected tidal radius of the 
satellite as a function of the distance to the point mass (Binney and 
Tremaine 1987); the mass ratio of the satellite and the main galaxy was 
taken to be 1/3 (this was done in accordance with our formal criterion 
for classifying a system as being of the M\,51 type (see the Introduction). 
As we see from the figure, in general, the satellites satisfy the tidal 
constraint imposed on their sizes. Moreover, the relatively more (circles) 
and less (diamonds) massive satellites show steeper and flatter dependences, 
respectively. If the satellite's size is limited by the tidal effect of 
the main component, then this is to be expected, because the tidal
radius is roughly proportional to (M$_s$/M$_m$)$^{1/3}$, where M$_s$/M$_m$
is the mass ratio of the satellite and the main galaxy 
(Binney and Tremaine 1987).

In ten of the twelve binary systems whose observations are presented 
in Klimanov et al. (2002), the satellite moves relative to the dynamical 
center of the main component in the same direction as the direction of 
rotation of the part of the main galaxy's disk facing it. In two 
cases (NGC\,2535/36 and  NGC\,4137), the motion of the satellite may be 
retrograde, although both main galaxies in these systems are seen almost
face-on and this conclusion is preliminary.
  
\begin{figure}
\centerline{\psfig{file=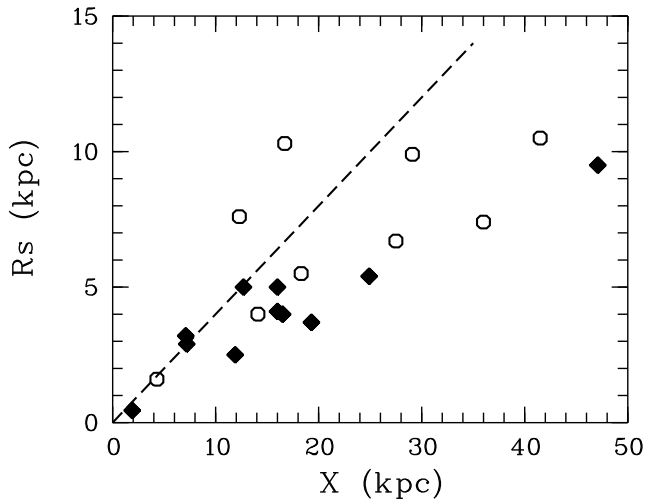,width=8.8cm,angle=-90,clip=}}
\caption{The relation between the observed optical radius of
the satellite and the projected linear distance to the main galaxy. 
The parameters of the satellites with $L_s/L_m>0.1$ and 
$L_s/L_m\leq0.1$ are indicated by the circles and diamonds, respectively. 
The straight line indicates the expected values of the tidal radius 
for a central point mass located (which approximates the main galaxy) 
and a mass ratio of the satellite and the main galaxy equal to 1/3.}
\end{figure}

\section{The Tully--Fisher relation}

The relation between luminosity and rotation velocity (the Tully--Fisher (TF) 
relation) is one of the most fundamental correlations for spiral galaxies.
This relation is widely used to study the large-scale spatial distribution 
of galaxies. In addition, it is an important test for the various kinds 
of models that describe the formation and evolution of spiral galaxies.

The observed rotation curves of M\,51-type galaxies are often highly 
irregular (see Fig. 1 in Klimanov et al. 2002). Therefore, to estimate 
the maximum rotation velocity, we used the empirical fit to the rotation 
curve suggested by Courteau (1997): 
\begin{center}
$v(r) = v_0 + v_c\,(1 + x)^{\beta} (1 + x^{\gamma})^{-1/\gamma}$, 
\end{center}
where $r$ is the distance from the dynamical center, $v_0$ is the 
line-of-sight velocity of the dynamical center, $v_c$ is the asymptotic 
rotation velocity, and $x = r_t/r$ ($\beta$, $\gamma$, and $r_t$ are 
the parameters). This formula well represents the rotation curves for 
most spiral galaxies (Courteau 1997).

Figure 4 shows the distribution of the parameters of M\,51-type galaxies 
in the absolute magnitude ($M(B)$)-maximum rotation velocity ($V_{max}$) 
plane. The values of $M(B)$ for the sample objects were corrected for 
internal extinction, as prescribed by the LEDA database. We took the 
values of $v_c$ (see above) obtained by fitting the observed rotation 
curves as the maximum rotation velocity for most objects. In
addition, these values were corrected in a standard way for the disk 
inclination to the line of sight and for the deviation of the spectrograph 
slit position from the galaxy's major axis. For eight galaxies, we 
estimated $V_{max}$ from the LEDA HI line widths.

\begin{figure}
\centerline{\psfig{file=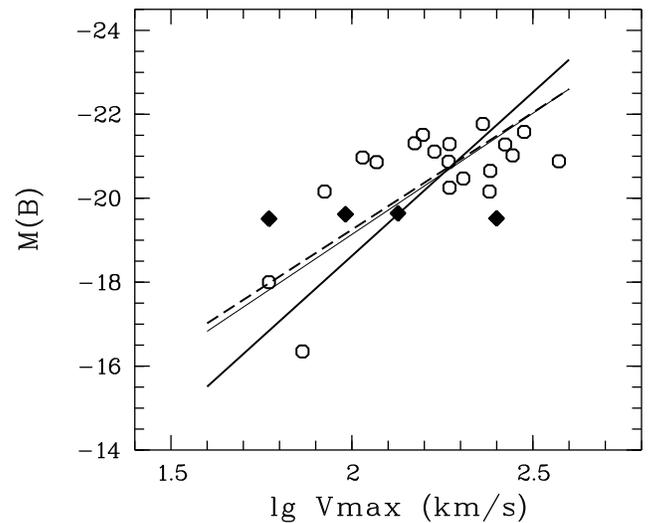,width=8.8cm,angle=-90,clip=}}
\caption{The Tully--Fisher relation for M\,51-type galaxies. The 
parameters of the main galaxies and their satellites are indicated 
by the circles and diamonds, respectively. The heavy solid straight 
line represents the TF relation for spiral galaxies as constructed 
by Tully et al. (1998) and the dashed line represents our TF relation 
for M\,51-type galaxies; the thin straight line indicates the TF relation 
for spiral galaxies at z$\sim$0.5 (Ziegler et al. 2002).}
\end{figure}

As we see from Fig. 4, the parameters of M\,51-type galaxies are located 
along a flatter relation than are those of nearby spiral galaxies. 
The mean relation for the objects of our sample indicated by the dashed
straight line is $L(B) \propto  V_{max}^{-(2.2\pm0.6)}$, while the relation
for local field spiral galaxies is $L(B) \propto V_{max}^{-(3.1\div3.2)}$
(Tully et al. 1998; Sakai et al. 2000). The bright (massive) binary members 
are located near the standard relation (Fig. 4). This location is 
consistent with the conclusion that the TF relation for giant spiral
galaxies does not depend on their spatial environment (Evstigneeva and 
Reshetnikov 2001). The M\,51-type galaxies with $V_{max} \leq 150$ km/s 
lie, on average, above the relation for single galaxies and, at a fixed
$V_{max}$, show a higher luminosity (or, conversely, at the same luminosity, 
they are characterized by, on average, lower observed values of $V_{max}$).

A similar fact (a different TF relation for the members of interacting 
galaxy systems and an excess luminosity of the low-mass members of 
these systems) was, probably, first pointed out by Reshetnikov (1994). 
The members of close pairs of galaxies also exhibit a flatter TF relation: 
$L(R) \propto V_{max}^{-2.2}$ (Barton et al. 2001). Barton et al. (2001) 
argue that interaction-triggered violent star formation in galaxies 
could be mainly responsible for the different slope of the TF relation 
for the binary members. Starbursts more strongly affect the observed 
luminosities of the low-mass galaxies by taking them away from the
standard TF relation.
	   
Violent star formation appears to be also responsible for the flatter 
TF relation for M\,51-type galaxies. As we showed previously (Klimanov 
and Reshetnikov 2001), IRAS data on the far-infrared radiation
from galaxies suggest an enhanced star formation rate in M\,51-type 
systems compared to local field objects. Unfortunately, the IRAS angular 
resolution is too low to separate the contributions from the main
galaxy and its satellite to the observed radiation. Figure 4 provides 
circumstantial evidence for violent star formation in both binary components.

Interestingly, a flatter TF relation has also been recently found 
(Ziegler et al. 2002) for spiral galaxies at redshifts $z\sim0.5$
(see Fig. 4). The similarity between the TF relations for nearby 
interacting/binary galaxies and distant spirals suggests that violent star
formation triggered by interaction (mergers) could be responsible for 
the observed luminosity evolution in distant low-mass galaxies. 
The rate of interactions and mergers between galaxies rapidly increases 
toward $z\sim1$ (Le Fevre et al. 2000; Reshetnikov 2000).
This mechanism must undoubtedly contribute to the luminosity evolution 
and, hence, to the observed TF relation.

\section{The evolution of the frequency of occurrence of 
M\,51-type galaxies}

Let us consider how the space density of M\,51-type galaxies changes 
with increasing redshift $z$. To this end, we studied the original frames 
of the Northern and Southern Hubble Deep Fields (Ferguson et al. 2000) 
and selected candidates for distant objects of this type. In selecting 
objects, we used the same criteria as those used to make the sample of
nearby galaxies. Unfortunately, there are no published spectroscopic $z$ 
estimates simultaneously for the main galaxy and for its satellite 
for any of the systems.

Each of the Deep Fields contains several thousand galaxies; the images 
of many of them are seen in projection closely or superimposed on one 
another, which makes it difficult to select candidates for M\,51-type
objects. Therefore, we restricted our analysis only to relatively bright 
and nearby galaxies ($z < 1.1$).

The candidates for distant M\,51-type objects are listed in the Table 1. 
The first column of the table gives the galaxy name from the catalog of 
Fernandez-Soto et al. (1999) (the first four and last three objects
are located in the Northern and Southern Fields, respectively); the 
second column gives the galaxy's apparent magnitude in the HST $I_{814}$ 
filter (for the first two systems, we took the estimates from the
catalog of Fernandez-Soto et al.; for the remaining systems, we provide 
our own magnitude estimates); the third column gives the redshift (the 
spectroscopic $z$ estimates (Cohen et al. 2000) are given with three 
significant figures; the photometric $z$ estimates (Fernandez-Soto et al. 
1999) are marked by a colon); the fourth column gives the spectral type,
which characterizes the galaxy's spectral energy distribution 
(Fernandez-Soto et al. 1999); and the last column contains our 
measurements of the angular separation between the nuclei of the main 
galaxy and its satellite.

\begin{table}[!ht]
\caption{Candidates for M\,51-type galaxies in the Hubble Deep
Fields}
\begin{center}
\begin{tabular}{|c|c|c|c|c|}
\hline
Name  & $I_{\rm 814}$ & $z$ & Spectral      &  $X$ ($''$) \\
      &               &     & type          &             \\ 
\hline
n45       & 21.26         & 1.012 & Scd            &   2.95 \\
n78       & 25.30         & 1.04: & Irr            &         \\
\hline
n350      & 21.27         & 0.320 & Scd            &   3.36 \\
n351      & 24.03         & 0.24: & Irr            &        \\
\hline
n888a     & 23.88         & 0.559 & Irr            &   0.72 \\
n888b     & 25.75         &       &                &        \\ 
\hline
n938a     & 23.11         & 0.557 & Scd            &   1.37 \\
n938b     & 25.85         &       &                &        \\
\hline
SB-WF-2033-3411a    &  22.76   & 0.55: & Irr       &  1.44   \\
\hspace*{2.65cm}b    &        24.94       &       &                &        \\
\hline
SB-WF-2736-0920a & 22.01  & 0.59: & Scd            &   1.54  \\
\hspace*{2.65cm}b   &          24.30    &       &                &        \\
\hline
SB-WF-2782-4400a & 23.05  & 0.53: & Irr            &   0.89  \\
\hspace*{2.65cm}b   &          24.58    &       &                &        \\
\hline
\end{tabular}
\end{center}
\end{table}

The integrated parameters of the selected candidates for M\,51-type 
systems are close, within the error limits, to those of nearby systems 
(see Section 2). Thus, the ratio of the observed luminosities
of the satellite and the main galaxy is 0.13$\pm$0.07 (in the $I_{814}$ 
filter, which roughly corresponds to the $B$ band in the frame of reference 
associated with the objects themselves at the mean redshift of the
sample under consideration); the mean observed separation between the 
main components and their satellites is 12$\pm$7 kpc. The absolute 
magnitude of the main galaxies that was estimated by applying
the $k$ correction (Lilly et al. 1995) is $M(B) =-19\fm4\pm1\fm4$. 
(These values, as well as those given below, were calculated for a 
cosmological model with a nonzero $\Lambda$ term:
$\Omega_m$=0.3, $\Omega_{\Lambda}$=0.7, H$_0$=65 km s$^{-1}$ Mpc$^{-1}$.)

To estimate the evolution rate of the space density of galaxies with $z$, 
we use the same approach that was used by Reshetnikov (2000). By 
assuming that the space density of M\,51-type galaxies changes as
\begin{center}
$n(z) = n_0(1 + z)^m$
\end{center}
($n(z)$ is the number of objects per unit volume (Mpc$^{-3}$) at redshift $z$ 
and $n_0 = n(z=0)$), we calculate the expected number of galaxies in the 
direction of the Deep Fields in the $z$ range concerned for various
exponents $m$.

The most important parameter required for this estimation is the 
spatial abundance of M\,51-type galaxies in a region of the Universe 
close to us ($n_0$). According to Klimanov (2003), M\,51-type galaxies 
account for 0.3\% of the field galaxies and about 4\% of the
binary galaxies. Interestingly, this estimate is close to the frequency 
of occurrence estimated by Vorontsov-Vel'yaminov (1978), who assumed 
that M\,51-type galaxies accounted for about 10\% of the interacting
galaxies. Assuming that about 5\% of the galaxies are members of interacting 
systems (Karachentsev and Makarov 1999), we find that M\,51-type systems
account for  0.5\% of all galaxies. By integrating the luminosity function 
of the field galaxies taken from the SDSS and 2dF surveys (Blanton et al.
2001; Norberg et al. 2002) in the range of absolute magnitudes $M(B)$ from 
-17$^m$ to -22\fm5, we can estimate the space density of field galaxies 
in this luminosity range as 0.017 Mpc$^{-3}$ and, hence, 
$n_0$ = 5.1\,10$^{-5}$ Mpc$^{-3}$.

Integrating the expression $n_0 (1 + z)^m$ over the $z$ range from 0.2 to 
1.1, we found that in the absence of density evolution ($m = 0$), the 
expected number of M\,51-type galaxies in the two fields is 0.8. The
observed number of objects (seven) exceeds their expected number by more 
than two standard Poisson deviations ($\sigma=\sqrt{7}=2.65$). Despite 
the poor statistics, this result provides evidence for the evolution of
the spatial abundance of M\,51-type objects with $z$. The exponent $m = 3.6$ 
corresponds to the observed number of galaxies. The formal spread in this 
value that corresponds to the range of the number of objects
from $7-\sqrt{7}$ to $7+\sqrt{7}$ is $^{+0.5}_{-0.8}$. The actual error in the
$m$ estimate can be much larger, for example, because of the uncertainty 
in $n_0$.

Of course, the estimated evolution rate depends on the assumed cosmological 
model. For example, for a flat Universe with a zero $\Lambda$ term and 
H$_0$=75 km s$^{-1}$ Mpc$^{-1}$, the exponent $m$ increases to 4.8.
      
\section{Discussion}

A typical M\,51-type binary system is a bright spiral galaxy with a 
relatively low-mass satellite physically associated with it located 
near the boundary of its disk. In most cases, the satellite shows prograde
motion; i.e., it rotates in the same direction as the main galaxy (see 
Section 2). Observational selection may be responsible for this peculiarity, 
because the satellite whose direction of orbital motion coincides
with the direction of rotation of the main galaxy can excite and maintain 
a large-scale two-arm pattern on the galactic disk. If the motion of the 
satellite is retrograde, then the tidal response will be much weaker. 
The presence of a spiral pattern is one of the criteria for classifying a 
galaxy as an M\,51-type object and three quarters of them show a large-scale
two-arm pattern (Klimanov and Reshetnikov 2001). Consequently, the predominance of prograde motions
of the satellites in objects with a well-developed two-arm spiral pattern 
can serve as a confirmation of the tidal nature of the spiral arms in 
such galaxies.

Other indications of the mutual influence of the galaxies in M\,51-type 
systems include an enhanced star formation rate (as evidenced by the high 
far-infrared luminosities of the systems (Klimanov and Reshetnikov 2001) 
and, possibly, by the flatter Tully--Fisher relation (Fig. 4)) and the 
existence of a tidal constraint on the sizes of the satellites (Fig. 3).

M\,51-type binary systems are relatively rare objects. According to 
Klimanov (2003), only  0.7\% of the spiral galaxies in the range of 
absolute $B$ magnitudes from --16$^m$ to --22$^m$ have a relatively bright
satellite near the end of the spiral arm and can be classified as objects 
of this type.

As was shown in the preceding section, the fraction of M\,51 type galaxies 
increases with redshift. Interestingly, within the error limits, the rate of
increase in the relative abundance of these objects ($m = 3.6^{+0.5}_{-0.8}$) 
roughly corresponds to the rate of increase in the number of binary and 
interacting galaxies. Thus, the fraction of galaxies with tidal
features is proportionalto $(1 + z)^m$ with $m=4\pm1$ (Reshetnikov 2000); 
the fractions of merging and binary galaxies evolve with 
$m = 3.4\pm0.6$ and $m = 2.7\pm0.6$, respectively (Le Fevre et al. 2000).

Let us now try to describe the possible origin and evolution of M\,51-type 
binary systems. Numerical calculations of the formation of galaxies performed
in terms of CDM (cold dark matter) models indicate that galaxies are formed 
within extended dark halos; many less massive subhalos must be contained 
within a massive halo (see, e.g., Kauffmann et al. 1993; Klypin et al. 1999). 
Thus, the halo of a galaxy with a mass comparable to the mass of the Milky Way 
must include several tens of satellites of different masses within its 
virial radius of 200--300 kpc (Benson et al. 2002a). Numerical calculations 
suggest that $\leq$5\% of the galaxies similar to the Milky Way have 
satellites with a $V$-band luminosity of --18$^m$ (Benson et al. 2002b). 
Consequently, such satellites (their luminosity is typical of the 
satellites of M\,51-type objects) are relatively but not extremely rare. 
Recall that our Galaxy has such a satellite, the Large Magellanic
Cloud, at a distance of about 50 kpc.

What is the subsequent fate of the relatively massive satellites formed 
in the halos of galaxies similar to the Milky Way? The evolution of the 
satellites will be governed mainly by two processes: (1) dynamical
friction, which will cause the satellite's orbital decay until the satellite 
merges with the main component; and (2) tidal stripping, which causes 
a decrease in the satellite's mass and, as a result, an increase in
the lifetime of its separate existence from the main galaxy. Recent 
studies suggest that the lifetime of a satellite under dynamical friction 
can be much longer than assumed previously (see, e.g., Colpi et al. 1999;
Hashimoto et al. 2003). In addition, this lifetime depends weakly on the 
orbital eccentricity of the satellite (Colpi et al. 1999).

Cosmological calculations indicate that the satellites formed within the 
massive halo of a central object are in highly elongated orbits with a mean 
eccentricity of $e = 0.6-0.8$ (Ghigna et al. 1998). If such satellites
are assumed to be observed in M\,51-type systems mostly near the orbital 
pericenter (in this case, their tidal effect on the disk of the main 
galaxy is at a maximum), i.e., $r_p=24$ kpc, then the distance at the
apocenter is 100--150 kpc and the orbital period of the satellite is 3--6 Gyr. 
In the Hubble time, such a satellite would make only 2 to 4 turns and 
might not be absorbed by the central galaxy by $z=0$. During its
first encounter with the main galaxy, the satellite can lose 50\% of its 
mass, which significantly increases the time of its orbital evolution 
(Colpi et al. 1999).
Consequently, the ``cosmological'' satellites formed on the periphery 
of the halos of central galaxies could in several cases survive by $z=0$ 
and be currently observed together with the main components as M\,51-type 
binary systems. 

Another, apparently more realistic scenario for the
formation of the binary systems under consideration is the capture of 
a relatively low-mass object by a central galaxy during their chance 
encounter. At present, the interactions of galaxies are relatively rare
events, but they are observed more and more often with increasing $z$ 
(at least up to $z\approx1$). Whereas only $\approx$5\% of the galaxies 
at $z=0$ are members of interacting systems with clear morphological evidence
of perturbation (Karachentsev and Makarov 1999), this fraction at $z=1$ 
is $\approx$50\% (Reshetnikov 2000; Le Fevre et al. 2000). Thus, an 
event during which a massive galaxy at $z\geq0.5$ captures a satellite seems
quite likely. This event is all the more likely since the
peculiar velocities of the galaxies were earlier lower and, hence, their 
chance encounters occurred with lower relative velocities and more often 
led to the formation of bound systems or mergers (see, e.g., Balland et al. 
1998). The orbital period in a circular orbit with a semimajor axis of 
$a=24$ kpc around a galaxy similar to the Milky Way is 0.5--1 Gyr. 
Consequently, in several Gyr (recall that an age equal to about half of
the Hubble time corresponds to $z=1$), most of such satellites will be 
absorbed by the main galaxies (Colpi et al. 1999; Penarrubia et al. 2002); 
by the time that corresponds to $z=0$, they must be observed rarely.
This scenario is confirmed by our evidence for the relatively rapid 
evolution of the space density of M\,51-type objects with $z$.

Thus, the currently observed M\,51-type systems could have both a 
primordial origin and a more recent origin -- through the capture of 
a satellite by the main galaxy. Mixed scenarios are also possible, for
example, when an encounter with another galaxy can change the orbit of 
the peripheral satellite and push it closer to the main galaxy. 
Of course, the actual pattern of formation and evolution of the 
binary systems under consideration is much more complex, and it
should be tested by numerical calculations.

How does the galaxy M\,51, the prototype of this class of objects, 
fit into the scenarios described above? It should be noted that in 
some respects, this binary system is not a typical representative of
the M\,51-type systems. For example, the mass ratio of the satellite 
and the main galaxy for it is 1/3 or even 1/2, a value that is larger 
than that for a typical binary system of this type. The dynamical 
structure of the binary system is not completely understood
either. The apparent morphology and kinematics of M\,51 can be explained 
both in terms of the models according to which we observe the separation 
of the galaxies after their first encounter (Durrell et al. 2003) and 
in terms of the approach according to which multiple encounters of the 
satellite and the main galaxy have already taken place in this system
(Salo and Laurikainen 2000). In the former case, the system M\,51 has, 
probably, formed recently (we observe it several hundred Myr after the 
passage of the satellite through the pericenter) during a close
encounter of the galaxies NGC 5194 and NGC 5195; in the latter case, this 
system is much older and its age can reach several Gyr.

In conclusion, note that because of the relatively small number of systems 
studied, some of our results (a different slope of the TF relation and an 
increase in the frequency of occurrence of M 51-type galaxies
with $z$) are preliminary and should be confirmed using more extensive 
observational data.

\bigskip
\section*{Acknowledgments}
{\it This study was supported by the Federal Program
``Astronomy'' (project no. 40.022.1.1.1101).}

\section*{REFERENCES}

\indent

1. Ch. Balland, J. Silk, and R. Schaeffer, Astrophys. J. 497, 541 (1998).

2. E. J. Barton, M. J. Geller, B. C. Bromley, et al., Astron. J. 121, 625 (2001).

3. A. J. Benson, C. G. Lacey, C. M. Baugh, et al., Mon. Not. R. Astron. Soc. 
333, 156 (2002a).

4. A. J. Benson, C. S. Frenk, C. G. Lacey, et al., Mon. Not. R. Astron. Soc. 
333, 177 (2002b).

5. J. Binney and S. Tremaine, Galactic Dynamics (Cambridge University Press, 
Princeton, 1987).

6. M. R. Blanton, J. Dalcanton, D. Eisenstein, et al., Astron. J. 121, 2358 (2001).

7. J. G. Cohen, D. W. Hogg, R. Blandford, et al., Astrophys. J. 538, 29 (2000).

8. M. Colpi, L. Mayer, and F. Governato, Astrophys. J. 525, 720 (1999).

9. S. Courteau, Astron. J. 114, 2402 (1997).

10. P. R. Durrell, J. Ch.Mihos, J. J. Feldmeier, et al., Astrophys. J. 582, 
170 (2003).

11. E. A. Evstigneeva and V. P. Reshetnikov, Astrofizika 44, 193, 2001.

12. H. C. Ferguson, M. Dickinson, and R. Williams, Ann. Rev. Astron. Astrophys. 
38, 667 (2000).

13. A. Fernandez-Soto, K. M. Lanzetta, and A. Yahil, Astrophys. J. 513, 34 (1999).

14. S. Ghigna, B. Moore, F. Governato, et al., Mon. Not. R. Astron. Soc. 300, 
146 (1998).

15. Y. Hashimoto, Y. Funato, and J. Makino, Astrophys. J. 582, 196 (2003).

16. I. D. Karachentsev, Binary Galaxies (Nauka, Moscow, 1987).

17. I. D. Karachentsev, D. I. Makarov, Galaxy Interactions at Low and High 
Redshift Ed. by J. E. Barnes and D. B. Sanders (1999), p. 109.

18. G. Kauffmann, S. D.M. White, and B. Guiderdoni, Mon. Not. R. Astron. Soc. 
264, 201, 1993.

19. S. A. Klimanov, Astrofizika 46, 191 (2003).

20. S. A. Klimanov and V. P. Reshetnikov, Astron. Astrophys. 378, 428 (2001).

21. S. A. Klimanov, V. P. Reshetnikov, and A. N. Burenkov, Pis'ma Astron. Zh. 
28, 643 (2002) [Astron. Lett. 28, 579 (2002)].

22. A. Klypin, A. V. Kravtsov, O. Valenzuela, and F. Prada, Astrophys. J. 522, 
82 (1999).

23. O. Le Fevre, R. Abraham, S. J. Lilly, et al., Mon. Not. R. Astron. Soc. 
311, 565 (2000).

24. S. J. Lilly, L. Tresse, F. Hammer, et al., Astrophys. J. 455, 108 (1995).

25. P. Norberg, Sh.Cole, C. M. Baugh, et al., Mon. Not. R. Astron. Soc. 336, 
907 (2002).

26. J. Penarrubia, P. Kroupa, and Ch.M. Boily, Mon. Not. R. Astron. Soc. 333, 
779 (2002).

27. V. P. Reshetnikov, Astrophys. Space. Sci. 211, 155 (1994).

28. V. P. Reshetnikov, Astron. Astrophys. 353, 92 (2000).

29. S. Sakai, J. R. Mould, S. M.G. Hughes, et al., Astrophys. J. 529, 698 (2000).

30. H. Salo and E. Laurikainen, Mon. Not. R. Astron. Soc. 319, 377 (2000).

31. R. B. Tully, M. J. Pierce, J.-Sh.Huang, et al., Astron. J. 115, 2264 (1998).

32. B. A. Vorontsov-Vel'yaminov, Astron. Zh. 52, 692 (1975) [Sov. Astron. Rep. 
19, 422 (1975)].

33. B. A. Vorontsov-Vel'yaminov, A. A. Krasnogorskaya, and V. P. Arkhipova, 
A Morphological Catalog of Galaxies (Izd. Mos. Gos. Univ., Moscow, 1962--1968), 
v. 1--4.

34. B.A. Vorontsov-Vel'yaminov, Exragalactic Astronomy (Nauka, Moscow, 1978).

35. B. L. Ziegler, A. Bohm, K. J. Fricker, et al., Astrophys. J. 564, L69 (2002).

\end{document}